\RequirePackage{ifpdf}
\documentclass{pCHEF}
\usepackage{arydshln}

\title{Irradiation Tests and Expected Performance of Readout Electronics of the ATLAS Hadronic Endcap Calorimeter for the HL-LHC}

\ShortTitle{ATLAS HEC Electronics for HL-LHC}

\author{\speaker{Martin Nagel}\thanks{on behalf of the ATLAS Liquid Argon
Calorimeter Group and the HECPAS Collaboration (IEP Ko\v{s}ice, Univ. of Montr\'{e}al, MPP Munich, IEAP Prague, NPI \v{R}e\v{z})}\\
        Max-Planck-Institut f\"{u}r Physik (Werner-Heisenberg-Institut)\\
        E-mail: \email{nagel@mpp.mpg.de}}

\abstract{The readout electronics of the ATLAS Hadronic Endcap Calorimeter (HEC) will have to withstand an about 3-5 times larger radiation environment at the future high-luminosity LHC (HL-LHC) compared to their design values. The preamplifier and summing boards (PSBs), which are equipped with GaAs ASICs and comprise the heart of the readout electronics, were irradiated with neutrons and protons with fluences surpassing several times ten years of operation of the HL-LHC. Neutron tests were performed at the NPI in Rez, Czech Republic, where a 36 MeV proton beam was directed on a thick heavy water target to produce neutrons. The proton irradiation was done with 200 MeV protons at the PROSCAN area of the Proton Irradiation Facility at the PSI in Villigen, Switzerland. In-situ measurements of S-parameters in both tests allow the evaluation of frequency dependent performance parameters, like gain and input impedance, as a function of fluence. The linearity of the ASIC response was measured directly in the neutron tests with a triangular input pulse of varying amplitude. The results obtained allow an estimation of the expected performance degradation of the HEC. For a possible replacement of the PSB chips, alternative technologies were investigated and exposed to similar neutron radiation levels. In particular, IHP 250 nm Si CMOS technology has turned out to show good performance and match the specifications required. The performance measurements of the current PSB devices, the expected performance degradations under HL-LHC conditions, and results from alternative technologies will be presented.}

\FullConference{Calorimetry for High Energy Frontiers - CHEF 2013,\\
		April 22-25, 2013\\
		Paris, France}

\usepackage{CHEF2013_definitions}
\begin{document}

\section{The ATLAS Hadronic Endcap Calorimeter}

The Hadronic Endcap Calorimeter (HEC) of the ATLAS experiment~\cite{JINST_3_S08003}
at the CERN Large Hadron Collider (LHC) is a copper-liquid argon
sampling calorimeter in a flat plate design~\cite{LArTDR,MPP-2007-237}, and provides coverage for hadronic showers in the
pseudorapidity range $1.5 < |\eta| < 3.2$. The HEC shares each of the
two liquid argon endcap cryostats with the electromagnetic endcap
(EMEC) and forward (FCAL) calorimeters, and consists of two wheels per
endcap. Each HEC wheel is made of 32 modules, each with 40 liquid argon
gaps, which are instrumented with active read-out pads. The signals
from the read-out pads are sent through short coaxial cables to
preamplifier and summing boards (PSB) mounted on the outer circumference of the
wheels inside the cryostat. The PSBs carry highly-integrated
preamplifier and summing amplifier chips in gallium-arsenide (GaAs)
technology. The signals from a set of preamplifiers are summed to one
output signal, which is transmitted to the cryostat
feed-through~\cite{MPP-2005-193}.

\section{Requirements of the HEC cold electronics for the HL-LHC upgrade}

The GaAs technology currently employed in the HEC cold electronics was selected for its excellent high frequency performance, stable operation at cryogenic temperatures, and radiation hardness~\cite{MPP-2005-193}. The radiation hardness specifications were defined for ten years of operation at the LHC design luminosity of $10^{34} \, \rm{cm^{-2} \, s^{-1}}$. The foreseen upgrade to the high-luminosity version of the LHC (HL-LHC) exceeds these limits by up to an order of magnitude~\cite{Atlas-Upgrade}. The expected radiation doses~\cite{ATL-GEN-2005-001} for an integrated luminosity of 3000 $\mathrm{fb}^{-1}$ in the region of the amplifier electronics located inside the endcap cryostats are $5.1 \cdot 10^{13} \, \mathrm{h} / \mathrm{cm}^{2}$ for hadrons above 20 MeV, $4.1 \cdot 10^{14} \, \mathrm{n_{eq}} / \mathrm{cm}^{2}$ for \mbox{1-MeV} equivalent neutrons in silicon (NIEL), and 6.2 kGy for the total ionizing dose (TID). These numbers include a total safety factor of 5 to account for uncertainties in the simulation. Proton and neutron irradiation tests have been performed to evaluate the limits of the current readout electronics and of potential alternative technologies under HL-LHC conditions~\cite{IEEEProc2012}.

\section{Tests}

The neutron irradiation tests were performed at the Fast Neutron Facility at the Nuclear Physics Institute of the ASCR in \v{R}e\v{z}, Czech Republic, up to an integrated fluence of ${2.2 \cdot 10^{16} \, \mathrm{n_{eq}^{(Si)}} / \rm{cm}^{2}}$. A $37\, \rm{MeV}$ proton beam incident on a $\rm{D_2O}$ target created a beam of neutrons with a continuous energy spectrum up to a maximum neutron energy of $36\, \rm{MeV}$ and a flux density up to $10^{11} \, \mathrm{n_{eq}^{(Si)}} / \rm{cm^{2}} / \rm{s}$. The proton irradiation test was performed at the Proton Irradiation Facility at the Paul Scherrer Institute in Villigen, Switzerland, with a monoenergetic beam of $200 \, \rm{MeV}$ protons up to an integrated fluence of $2.6 \cdot 10^{14} \, \mathrm{p} / \rm{cm}^{2}$. The transistors were bonded in ceramic casings and mounted on boards, which were placed in an aluminum frame with 17 mm distance between slots and aligned in the particle beams. The three different transistor technologies being tested were Si CMOS FET in SGB25V 250nm technology from IHP, SiGe Bipolar HBT (IHP SGB25V 250nm and IBM 8WLBiCMOS 130nm), and the GaAs FET currently used in ATLAS, either the Triquint CFH800 250nm transistors themselves or integrated into the HEC BB96 Preamplifiers (PAs) and Systems. The performance of the transistors was monitored \textsl{in situ} during irradiation at room temperature with a vector network analyzer recording a full set of S-parameters~\cite{SParameter}, which were converted to standard transistor parameters using suitable small signal circuit models. In addition, the neutron irradiated boards were re-tested 3 months after the neutron test at the Max-Planck-Institute for Physics (MPP) in Munich, immersed in liquid nitrogen to simulate the conditions inside the HEC cryostats.

\section{Flux determination}

A combination of radiation films and radiation monitoring diodes, placed at various slot positions along the beam, were used to determine the beam profile and, together with the beam current measurements, the particle flux. The films were exposed to a given beam current for various amounts of time and subsequently scanned. An iterative calibration procedure was used to obtain the dose as a function of net optical density~\cite{IEEEProc2012,RadfilmCalib}, by combining the known relative dose measurements with the absolute normalization obtained from MC simulations and from the diodes. For the neutron test with its continuous neutron energy distribution, the fluence was scaled to the \mbox{1-MeV} equivalent neutron fluence in silicon, employing the non-ionizing energy loss (NIEL) scaling hypothesis~\cite{NIEL}, independent of the actual technology under irradiation.

\section{Results}

\begin{figure}[htb]
  \begin{center}
    \includegraphics[width=0.49\textwidth]{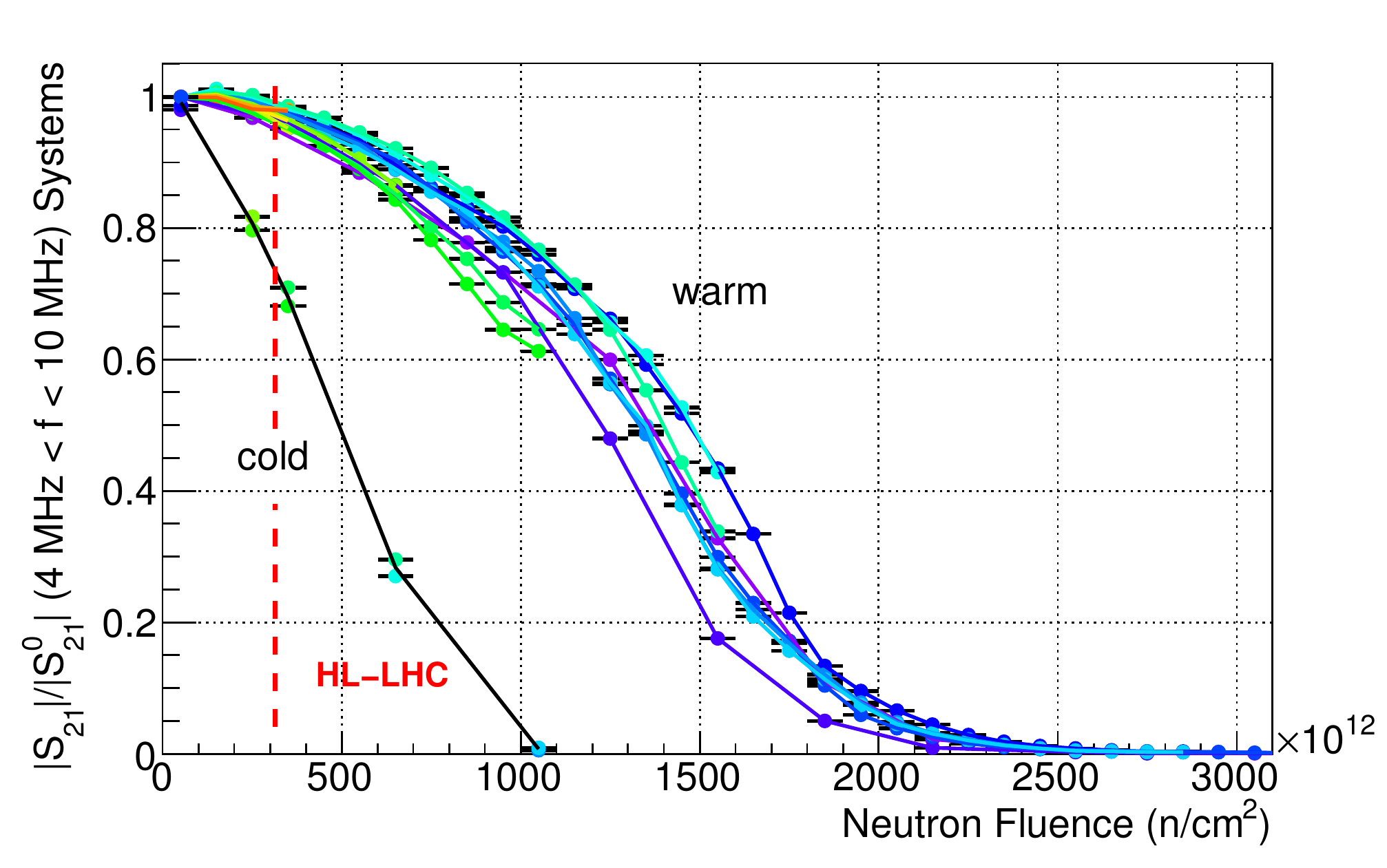}
    \includegraphics[width=0.49\textwidth]{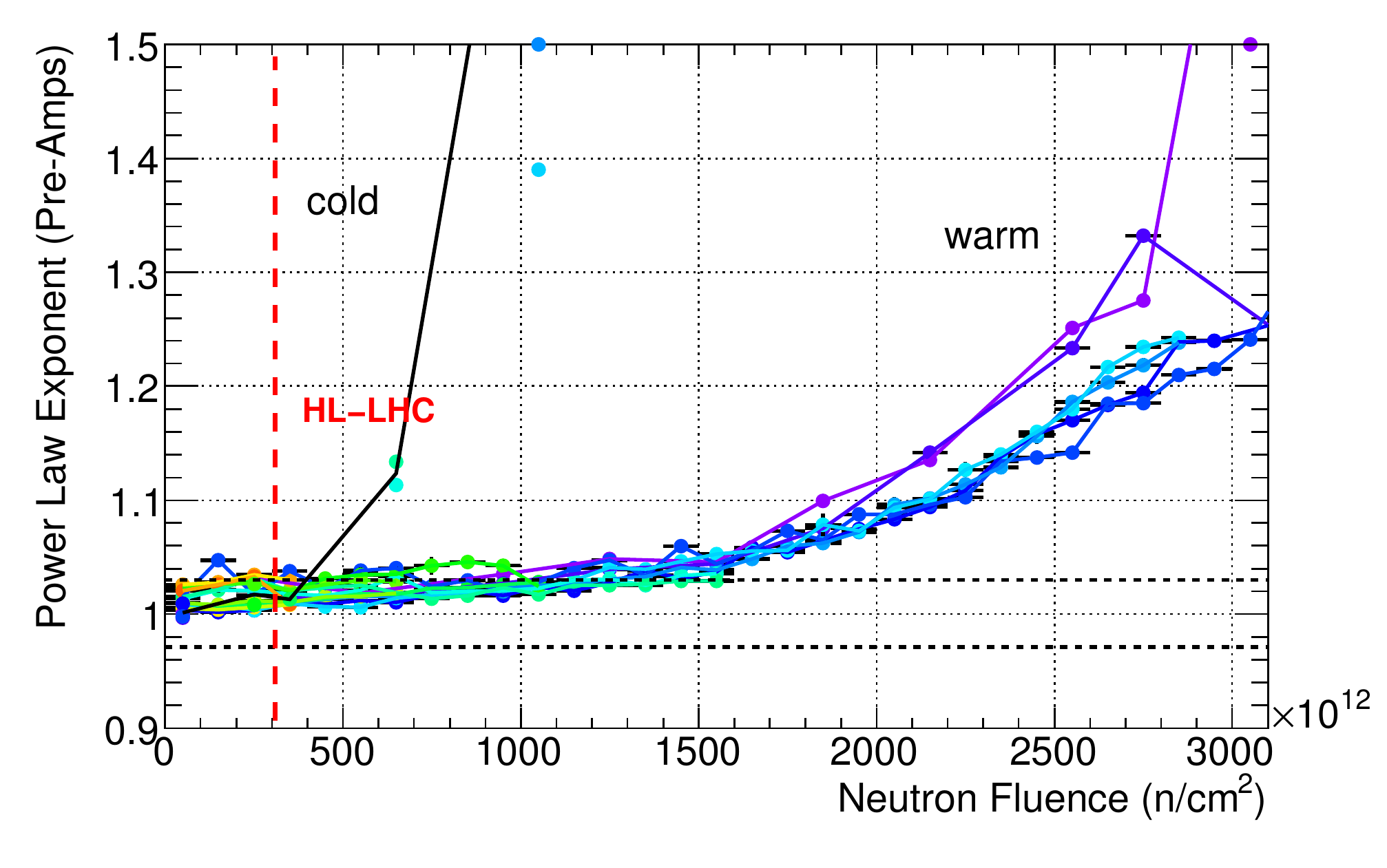}
    \caption{Forward transmission coefficient $|S_{21}|$ of the BB96 Systems normalized to the value before irradiation (left), and linearity of the BB96 PAs in terms of a power-law exponent (right). Dense data points connected by colored lines correspond to the \textsl{in situ} measurements at room temperature at NPI, whereas the black line represents the cold results obtained at MPP. The red dashed line indicates the HL-LHC limit (including the safety factor of 5), taking into account that $4.1 \cdot 10^{14} \, \mathrm{n_{eq}^{(Si)}} / \mathrm{cm}^{2}$ NIEL in ATLAS corresponds to $3.0 \cdot 10^{14} \, \mathrm{n_{eq}^{(Si)}} / \mathrm{cm}^{2}$ NIEL at NPI (for GaAs devices)~\cite{IEEEProc2012}.}
    \label{BB96_gain_lin}
  \end{center}
\end{figure}

Figure~\ref{BB96_gain_lin} shows two performance parameters of the currently installed BB96 ASICs, the forward transmission coefficient of the Systems and the linearity of the PAs, as a function of neutron fluence. The linearity of the PAs was characterized by a power-law fit to their response to triangular input pulses of varying amplitudes. The warm results show only a small degradation of the HEC electronics up to HL-LHC conditions, whereas the cold measurements indicate a much more severe performance degradation.

The impact of the electronics degradation under HL-LHC conditions on the HEC performance has been studied using simulated di-jet events and including both gain losses and non-linearity effects. The full HEC electronics chain has been simulated, with the appropriate radiation damage of the electronics applied cell-by-cell and layer-by-layer before digitization and reconstruction. The HEC calibration system can compensate on average for gain losses, but the non-linearities of the individual PAs, which are summed passively by the driver stage of the BB96 Systems, cannot be corrected by means of the calibration system. Preliminary results of this ongoing effort indicate a significant deterioration of the jet energy resolution (up to 50\%) and linearity (about 10\%).

\begin{figure}[htb]
  \begin{center}
    \includegraphics[width=0.49\textwidth]{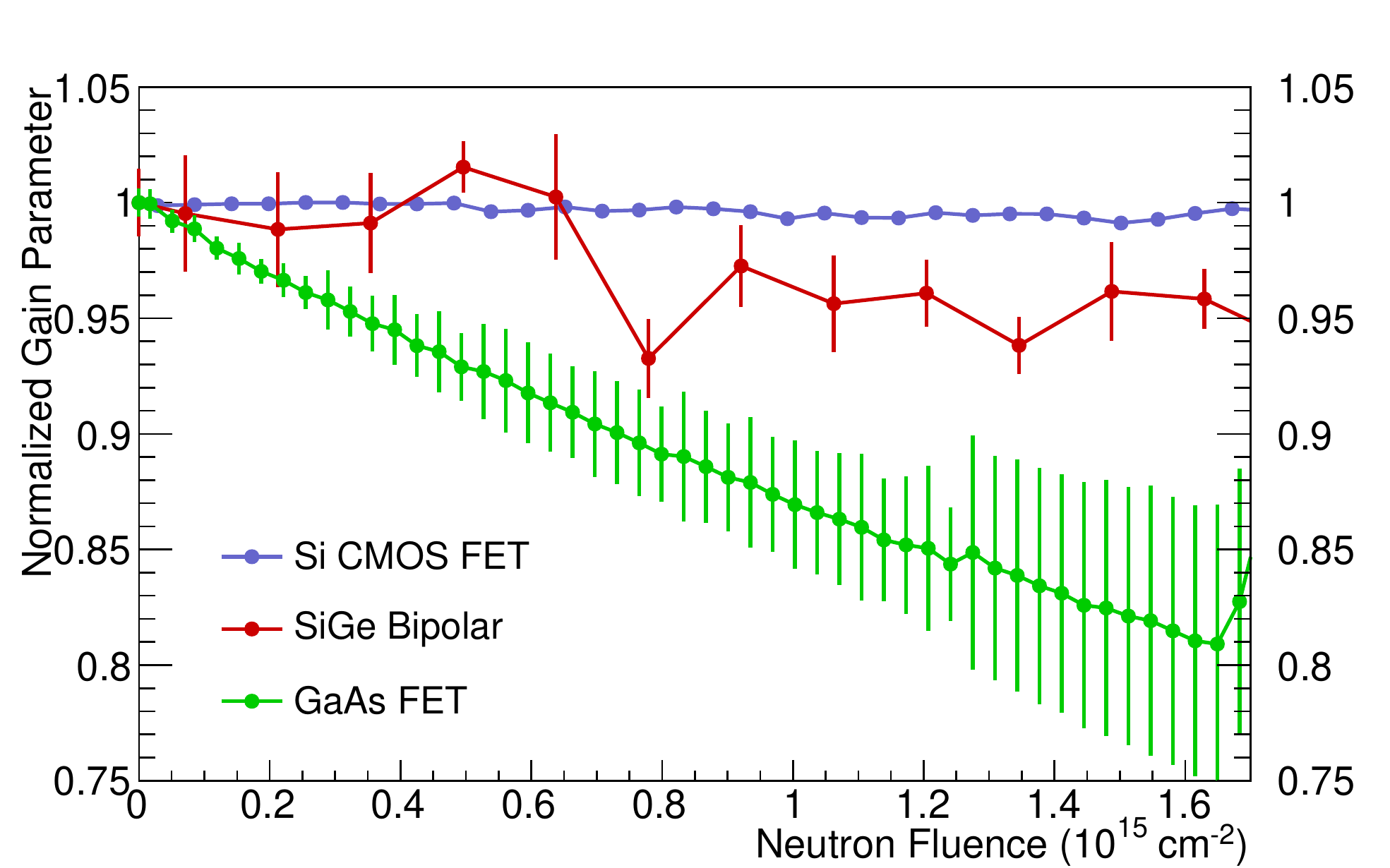}
    \includegraphics[width=0.49\textwidth]{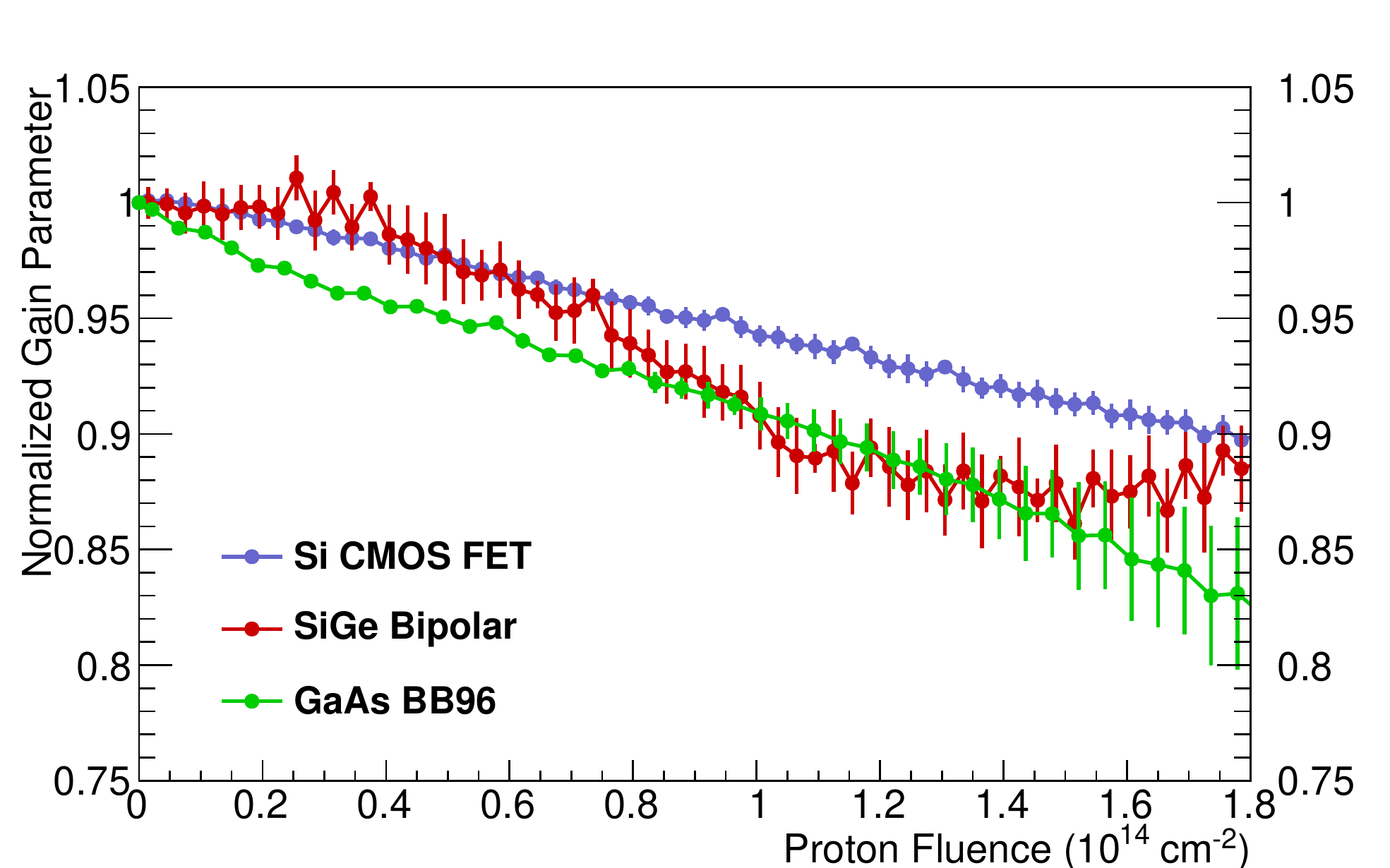}
    \caption{Relative gain loss as a function of proton (left) and neutron (neutron) fluence for the various technologies described in the legend; the respective gain parameters are described in the text. The data points represent the average of all the devices of a particular technology that were irradiated up to a given particle fluence; the error bars are the sample RMS. The neutron fluence is measured as 1-MeV equivalent Si NIEL~\cite{5402221}.}
    \label{Trans_gain}
  \end{center}
\end{figure}

The results of the investigation of alternative technologies are summarized in Fig.~\ref{Trans_gain}, which shows the two beam test results in terms of the appropriate gain parameters as a function of the corresponding particle fluence, normalized to the corresponding value before irradiation. The various gain parameters are the real part of the transconductance $g_m$ for the FET transistors, the differential current gain $\beta$ for the bipolar transistors, and in case of the HEC BB96 PAs and Systems, where our simple transistor models are not applicable, the transresistance $r_m = \frac{\Delta V_{\rm out}}{\Delta I_{\rm in}} = |S_{21} \cdot Z_{\rm in}|$, which relates the output voltage to the input current, where $Z_{\rm in}$ is the input impedance. Displayed are the sample averages of all the devices of a particular technology under test at each of the two test facilities, that were irradiated up to a given particle fluence; the error bars represent the sample RMS. The relative change of the gain parameter is quoted in Table~\ref{Result_table}, where a linear extrapolation of the gain parameter has been used for those devices not reaching the required limits.

\begin{table}[htb]
  \begin{center}
    \begin{tabular}{|l|c|c:c|}
      \cline{2-4}
      \multicolumn{1}{c|}{} & Neutron test & \multicolumn{2}{|c|}{Proton test$^{\dagger}$}\\
      \hline
      Technology & NIEL: $4.1 \cdot 10^{14} \, \rm{n_{eq}^{(Si)}/cm^2}$ & \multicolumn{1}{|c:}{$\Phi_{\rm had}$: $5.1 \cdot 10^{13} \, \rm{h/cm^2}$} & TID: $6.2 \, \rm{kGy}$\\
      \hline
      Si CMOS FET & 0\% & -3\% & -1\%\\
      SiGe Bipolar & -1\% & -3\% & -1\%\\
      GaAs FET & -5\% & n/a & n/a\\
      GaAs BB96 (warm) & -8\% & -5\% & -1\%\\
      GaAs BB96 (cold)$^{\dagger}$ & -18\% & n/a & n/a\\
      \hline
    \end{tabular}
    \caption{Loss of gain of various transistor technologies under neutron and
    proton irradiation~\cite[$^{\dagger}$7]{5402221}.}
    \label{Result_table}
  \end{center}
\end{table}

\section{Summary}

The GaAs technology currently used in the HEC cold electronics degrades measurably at the expected radiation levels, in particular under the more realistic cold conditions. The gain losses can be calibrated out on average, but the non-linearities of the PAs cannot be corrected. The two effects taken together lead to a significant degradation of the HEC performance, which needs further investigation in terms of its impact on physics measurements.

Regarding alternative technologies, both the Si CMOS FETs and the SiGe bipolar transistors are more radiation hard than the currently used GaAs FETs. Preference is given to the Si CMOS FETs, since the SiGe Bipolar transistors, due to their operation at cold temperatures, require a stabilization of their operation point~\cite{5402221}, which would lead to a more complex circuitry.

\bibliography{CHEF2013_bib_Nagel}{}
\bibliographystyle{unsrt}

\end{document}